\documentstyle[11pt]{article}
\voffset = - 1.7 cm
\hoffset = - 1.3 cm
\newcommand{\be}{\begin{equation}}
\newcommand{\ee}{\end{equation}}
\newcommand{\bea}{\begin{eqnarray}}
\newcommand{\eea}{\end{eqnarray}}
\newcommand{\nn}{\nonumber \\}
\newcommand{\p}{\partial}
 
\begin{document}
\rightline{\Large CGPG-97/2-1}
\vskip .6cm 
\centerline{\LARGE Gravity and BF theory defined in bounded regions}
\rm
\vskip .6cm
\centerline{Viqar Husain${}^\dagger$ and Seth Major${}^*$}
\vskip .3cm
\centerline{\it  Center for Gravitational Physics and Geometry} 
\centerline{\it Department of Physics, The Pennsylvania State 
University}
\centerline{\it University Park, PA, USA 16802}
\vskip .3cm
\centerline{February 12, 1997}
\vskip .3 cm
\centerline{ABSTRACT}

We study Einstein gravity in a finite spatial region.  By requiring a
well-defined variational principle, we identify all local boundary
conditions, derive surface observables, and compute their algebra.
The observables arise as induced surface terms, which contribute to a
non-vanishing Hamiltonian. Unlike the asymptotically flat case, we
find that there are an infinite number of surface observables. We give
a similar analysis for $SU(2)$ BF theory.

\noindent
\vskip .3cm
internet addresses:  ${}^*$seth@phys.psu.edu 
${}^\dagger$husain@phys.psu.edu

\section{Introduction}

Gravity has been studied mainly in the context of either closed or
asymptotically flat spacetimes. The former applies to cosmology,
whereas the latter applies to situations where the gravitating system
is viewed from a flat environment at infinity. The asymptotically flat
setting allows the identification of properties of the system, such as
its energy and angular momentum \cite{rt,bm,ah}.  In the case of
asymptotically flat spacetimes, these conserved quantities, being
integrals over a two-sphere, may be called ``surface observables''.

There exist, however, interesting solutions of the Einstein equations
which do not fall into the closed or asymptotically flat category.
These require a study of more general boundary conditions.  Recently
such boundary conditions have been studied. Brown and York \cite{BY}
study the covariant gravity action for a spatially bounded region, and
derive surface observables on a finite boundary (which are refered to
as quasi-local quantities). Balachandran, Chandar and Momen \cite{BCM}
perform a similar analysis in the context of an inner boundary and an
asymptotically flat outer boundary. Hawking and Horowitz \cite {HH},
provide an analysis for asymptotic conditions other than the 
asymptotically flat one. The main approach underlying all these works is
functional differentiability of the gravity action in the presence of
boundaries, as in the initial work of Regge and Teitelboim \cite{rt}.

The study of more general boundary conditions falls into two main
categories: A gravitating system may be viewed as being enclosed in a
finite spatial region, or, as in the asymptotically flat situation, in
a 2-sphere at infinity. In this paper we study the former case with an
emphasis on presenting {\it all} possible boundary conditions, and
finding observables of the theory.

Observables for gravity, or any other generally covariant field
theory, may be defined as phase space functionals that commute weakly
with the first class constraints of the theory. For a four-dimensional
theory in a finite spatial region, observables may be classified into
``bulk'' and ``surface'' observables. The former are integrals over
the spatial region, while the latter are integrals over the surface
bounding the spatial region.

There are (at least) two reasons why it may be useful to find surface
observables and their algebra for a gravitating system in a bounded
spatial region.  The first reason has to do with black hole
entropy. Specifically the questions are: What are the microscopic
degrees of freedom of a black hole? Where do these degrees of freedom
reside?

Recently there has been a proposal, originating in string theory, for
the statistical mechanical interpretation of black hole entropy
\cite{SV}. In the weak coupling limit of string theory, there are
bound states of D-branes labeled by charges {\it which are the same}
as the charges on the extremal black holes.  The degeneracy of these
bound states is taken to represent the microscopic degrees of freedom
of the black holes -- which arise only in the strong coupling limit.
It is remarkable that this degeneracy leads to the correct entropy
formula for black holes. However, essential to the identification of
these states as black hole microstates is the extrapolation of the
degeneracy calculation from weak to strong coupling (known as the
``non-renormalization theorem'').  This extrapolation obscures the
spacetime origin of the microscopic degrees of freedom in the strong
coupling limit (where there is a black hole), as well as the location
of the degrees of freedom.  Furthermore, as this idea applies only to
extremal and near extremal black holes, it does not work for the
Schwarzschild black hole. Therefore this string theory approach so far
provides only an indirect answer to the two questions.

Another conjectured solution, investigated in detail by Carlip
\cite{SC} for a black hole in ($2+1$)-dimensions \cite{BTZ}, provides
the following answer to these questions: The microscopic degrees of
freedom of a black hole are those of a theory induced on the horizon.
This horizon forms the (null) boundary of the system.  ``Surface
observables'' for the whole system are observables of the induced
boundary theory.  The answer arises by first noticing that
($2+1$)-gravity with a cosmological constant may be expressed as a
Chern-Simons theory \cite{at,witten}.  This theory, on a manifold with
boundary, induces the two-dimensional WZNW theory on the
boundary. Since ($2+1$)-gravity has a finite number of degrees of
freedom, and the WZNW theory has an infinite number, this effect of
inducing the WZNW theory on the boundary is referred to as the ``bulk
gauge degrees of freedom becoming dynamical on the boundary.'' The
conserved currents of this theory form a Kac-Moody algebra, as do the
surface observables. Quantization of the surface observable algebra
gives a Hilbert space of states associated with the boundary, from
which the entropy is determined. It is not clear whether this approach
will work for ($3+1$)-gravity.

The second reason for a full investigation of surface observables is
the ``holographic hypothesis'' \cite{t'hooft,suss}. This hypothesis
rests on the assumption that the maximum allowed entropy in a region
bounded by a spherical surface of area $A$ is $A/4$, corresponding to
a black hole that just fits in the surface. This finite entropy
implies a phase space of finite volume, and hence a finite dimensional
Hilbert space for the system. 't Hooft \cite{t'hooft} further argued
that this leads to the striking conclusion that physical degrees of
freedom must be associated with the boundary of the region: If the
entropy of a bounded system not containing a black hole were
proportional to its volume, then one could add matter until the system
becomes a black hole and the entropy becomes proportional to the
area. The entropy would decrease in such a process, and lead to an
apparent violation of the second law of thermodynamics. One solution
to this conundrum is to hypothesize that the entropy of a bounded
system must always be proportional to the boundary area.  This follows
if the degrees of freedom are associated only with the boundary.

It may be possible to verify this hypothesis using the present work,
if one can quantize the algebra of boundary observables such that the
resulting representation space has finite dimension. If the observable
algebra is infinite dimensional this may not be possible unless only a
finite, and somehow ``representative'' subset is quantized.  In the
context of canonical gravity and specific boundary conditions, a
quantization of a set of surface observables, including an area
observable, has been studied recently \cite{lee}.

To investigate these issues, we provide an analysis for boundary
conditions at a finite spatial boundary for Einstein gravity, and also
for the topological BF theory in four dimensions.  In two specific
cases, we exhibit the surface observables of the system, and compute
their algebra. There is earlier work on the finite boundary case for 
gravity  in \cite{BY} and \cite{SL}, and for Abelian BF theory in 
\cite{bal}.
 
We begin in the next sub-section by reviewing the derivation of the
ADM surface observables for asymptotically flat general
relativity. Following this is a brief discussion of the spatially
closed case in the Ashtekar Hamiltonian formulation.  This sets the
stage for our discussion of Einstein gravity in a bounded spatial
region. In Section II we give a general procedure for constructing
surface observables, followed by a discussion of the possible local
boundary conditions and corresponding surface observables for general
relativity in a finite spatial region. Section III contains a similar
analysis for the topological $SU(2)$ BF theory. The generalization of
the results to gauge groups other than $SU(2)$ is immediate. The final
section presents our conclusions and contains a comparison with other
studies of gravity in a finite spatial region.
 
\subsection{The asymptotically flat case: A brief review}

The fundamental difference between the variational principles for the
Einstein equations for spatially open versus closed spacetimes is that
the former requires proper treatment of asymptotic boundary
conditions; surface terms need to be added to the
action\footnote{Here, we take the view that the action { \em is} the
action for the compact case.}  to make the variational principle well
defined.  One crafts these surface terms so that their variation
cancels the surface terms arising from the variation of the ``bulk''
part of the action.

The standard asymptotically flat spacetime
is defined so that for large proper radial coordinate $r$
(spatial infinity), the spacetime metric behaves like the
Schwarzschild metric 
\be 
ds^2|_{r \rightarrow \infty} = - \left( 1-
{2M\over r } \right) dt^2 + \left( \delta_{ab} + M {x^a x^b\over r^3}
\right) dx^a dx^b, 
\ee 
where $\delta_{ab}$ is the Euclidean 3-metric and $x^a$ are the asymptotic
cartesian coordinates.  The definition of the phase space for
asymptotically flat spacetimes includes specific fall-off conditions
for the spatial metric $q_{ab}(x,t)$, its conjugate momentum
$\pi^{ab}(x,t)$, and for the lapse and shift functions $N(x,t)$ and
$N^a(x,t)$.  These conditions completely determine the allowed gauge
symmetries. In the asymptotically flat case, spacetime diffeomorphisms
are restricted to Poincar\'e transformations in the asymptotic
(Minkowski) region. The lapse and shift functions have the asymptotic
form
\bea
\label{shiftFO}
N &\rightarrow& \alpha + \delta_{ab}\beta^b x^a + O(1/r)\nn
N^a &\rightarrow& \alpha^a + \epsilon^a_{\ bc}\phi^b x^c + O(1/r), 
\eea
where $\alpha$ and $\alpha^a$ are time and space translations,
and $\beta^a$ and $\phi^a$ are boost and spatial rotation parameters. 
The fall-off conditions on the metric and its conjugate momenta, guided 
by the Schwarzschild metric, are 
\bea 
q_{ab}&\rightarrow& \delta_{ab} + {f_{ab}(\theta,\phi)\over r} + 
O(1/r^2)\nn
\pi^{ab}&\rightarrow& {p^{ab}(\theta,\phi)\over r^2} + O(1/r^3). 
\eea 
The fall off conditions, by themselves, are not sufficient to make the
Hamilton variational principle well defined; one must add boundary
terms to the action \cite{rt}.  With these choices, together with
parity conditions on the angle dependent tensors $f_{ab}$ and
$p^{ab}$, the surface terms that need to be added to the (3+1)-action
for the compact case are precisely the ADM four-momentum, angular
momentum and boost charge \cite{rt}.

Functional differentiability of the (3+1)-action, or equivalently, the
constraints, requires surface terms to be added to the action.  The
full Hamiltonian becomes a linear combination of constraints {\it
plus} surface terms. Evaluated on a classical solution, the initial
value constraints vanish leaving a non-vanishing ``surface
Hamiltonian''. This Hamiltonian is the sum of conserved charges
corresponding to the lapse and shift functions in the asymptotic
region.

While the full Hamiltonian is functionally differentiable one still
has to check that its Poisson algebra closes. The algebra does close
and reduces to the Poincar\'e algebra when evaluated on a
solution. Thus, the algebra of the full Hamiltonian with itself {\em
necessarily} gives the surface observable algebra.

We can then ask if there are surface observables {\it other} than
those which are already included in the full Hamiltonian. Are there
other phase space functionals associated to the boundary which commute
with the full Hamiltonian? An immediate attempt might be to see if
more freedom can be introduced into the lapse and shift functions
which parameterize the ADM surface observables.  As an example
consider the candidate observable defined using the diffeomorphism
generator
\begin{displaymath}
\int_\Sigma d^3x\ 
\pi^{ab} { \cal L}_M g_{ab} \approx -2 \int_{\p\Sigma} d^2S_a\
M_b\pi^{ab}, 
\end{displaymath}
where the vector field $M^a$ is now arbitrary. The Poisson
bracket of this functional with the full Hamiltonian gives a
non-vanishing surface term {\it unless} $M^a\rightarrow \epsilon^a_{\
bc}\phi^b x^c$; that is, it is an observable only if it reduces to the
familiar ADM angular momentum. One can check similarly that no new
surface observables arise  using the Hamiltonian
constraint.\footnote{There is more freedom in the boundary
observables than has been manifested so far \cite{ah}. The lapse and
shift functions of Eq. (\ref{shiftFO}) can have additional angle
dependent functions.  These are the so called ``super translations,''
which are transformations on the 2-sphere at infinity; these are in 
addition to the translations, rotations and boosts already present in 
Eq. (\ref{shiftFO}).}

Intuitively, one expects a connection between the freedom in the lapse
and shift at the boundary, and the number of surface observables: A
reduction in the number of gauge transformations at a boundary should
correspond to an increase in the number of surface observables.  As we
will see, this expectation only partially true.
  
For gravity in a bounded region, to be discussed in Section II, we
follow a method similar to the one used above.  While we do not work
with an asymptotic region, with its corresponding forms for the lapse,
shift and phase space variables, there are nonetheless significant
restrictions on the boundary variables. These form the possible
boundary conditions for gravity.  Before proceeding to this, we first
review the canonical theory on a compact manifold in the Ashtekar
variables.

\subsection{The compact case: Constraints and algebra}

For spacetime manifolds ${ \cal M }=\Sigma \times R$, where $\Sigma$ is
closed, the (3+1)-action for vacuum, Riemannian general relativity in
the Ashtekar variables is
\be
S[E^{ai}, A_a^i; \Lambda^i, N^a, N]  = { 1 \over \kappa } 
\int_{t_1}^{t_2} dt \int_\Sigma d^3x\  
[\ E^{ai}\dot{A}_a^i - N{\cal H} - N^a{\cal D}_a 
-\Lambda^i G^i\ ].
\label{3+1}
\ee
The conjugate phase space variables are an $SU(2)$ connection $A_a^i$
and a densitized, inverse triad $E^{ai}$, which satisfy
\be
\{ A_a^i(x), E^{b}_j(y) \} = \kappa \, \delta_a^b \delta^i_j 
\delta(x,y), 
\ee
with $\kappa = 16 \pi G$. (We set $\kappa = 1$ in the following.)
The Lagrange multipliers $N$, $N^a$, and
$\Lambda^i$ are, respectively, the lapse, shift, and $SU(2)$ gauge
rotation parameters. Varying the action with respect to these
functions gives the first class constraints
\bea
{\cal G}^i &\equiv& - D_aE^{ai} \approx 0;
\label{gauss} \\ 
{\cal D}_a &\equiv&  - E^{bi}\partial_a A_b^i + \partial_b(E^{bi}A_a^i) 
\approx 0; 
\label{diffeo} \\
 {\cal H} &\equiv& - \epsilon^{ijk}F^k_{ab}E^{ai}E^{bj} \approx 0, 
\label{hamiltonian} 
\eea
where  $D_a \lambda^i = \partial_a \lambda^i + \epsilon^{ijk}
A_a^j \lambda^k$ and 
$F_{ab}^i = \p_a A_b^i - \p_b A_a^i + \epsilon^{ijk}A_a^j A_b^k$.
These constraints generate gauge transformations via the Poisson
bracket. The smeared diffeomorphism constraint
\footnote{The theory originally found by a Legendre transform 
\cite{AA} 
has the vector constraint
$$
V(N) = \int_\Sigma d^3x\ N^a E^{bi} F_{ab}^i.
$$
However, the Gauss constraint may be combined with this to give
the diffeomorphisms constraint
$$
D(M) = V(M) - G(A_a^iM^a) =
\int d^3x\ M^a\ [\ E^{bi}\partial_a A_b^i - \partial_b(E^{bi}A_a^i)\ ]
$$
used above.}
$D(N)=-\int d^3x\ N^a{\cal D}_a$ satisfies the expected relations 
\begin{equation}
\{ A_a^i, D(N) \} = { \cal L}_N A_a^i, \ \ \ \ \{ E^{ai}, 
D(N) \} = { \cal L}_N E^{ai}.
\end{equation}
As the manifold has no boundary, there is some freedom in writing the
constraints.  For instance, integrating the second term of the
diffeomorphism constraint (Eq. (\ref{diffeo})) by parts one finds,
\be
D(N) = \int_\Sigma d^3x\ E^{ai} { \cal L }_N A_a^i.
\ee

The Hamiltonian constraint has density weight $+2$ so the
lapse function has density weight $-1$.  The resulting constraint
\be
H(N) = \int_\Sigma d^3x\ N \epsilon^{ijk} E^{ai} E^{bj} F_{ab}^k
\label{ham}
\ee
generates time evolution via the Poisson bracket.

Classically, the constraints satisfy the following algebra \cite{AA}
\begin{eqnarray}
\left\{ G(N), G(M) \right\} & = & - G([N,M]) \label{cont1} \\
\left\{ D(N), G(M) \right\} & = & - G( {\cal L}_N M) \\
\left\{ D(N), D (M) \right\} & = & - D ([N,M]) \\
\left\{ G(N), H(M) \right\} & = & 0 \\
\left\{ D(N), H(M) \right\} & = & - H( {\cal L}_N M) \\
\left\{ H(N), H(M) \right\} & = & D(K) + G(A_a K^a) \label{cont2}
\end{eqnarray}
where $K^a = E^{ai} E^{bi} \left( N \partial_b M - M \partial_b N
\right)$.

A more complete discussion of the spatially closed case can be found
in Ref. \cite{AA}.  The asymptotically flat case is presented in
Refs. \cite{AA,TT}.  In the next section we consider the possible 
boundary conditions for gravity in a finite spatial region.

\section{Gravity in a bounded spatial region}

We consider spatial slices $\Sigma$ with boundary $\p\Sigma$. The
boundaries are taken to be ``orthogonal'' in the sense that the normal
$n_a$ to the spatial boundary is orthogonal to the timelike direction
of the foliation. (This condition does not rule out asymptotic
boundaries or bifurcate horizons.)  Though our discussion focuses on a
single boundary, the analysis can be extended easily to a boundary
with disjoint regions.  In this case one can choose separate
boundary conditions and surface terms for each disjoint region of the
boundary. Our analysis proceeds in the following steps: 

\begin{description}

\item{(1)} When a boundary is present the variations of the
(3+1)-action (\ref{3+1}) with respect to the phase space variables
$E^{ai}$ and $A_a^i$ are not defined.  To define the theory, one must
add appropriate surface terms to the action and impose boundary
conditions.  There are a number of ways to do this, and we list all 
the possible choices.

\item{(2)} We find the full Hamiltonian $H_F$ (constraints plus
surface terms), which is a function of all gauge parameters, and
compute the algebra of $H_F$ with itself. This Poisson bracket should
close in the same way that the constraint algebra closes; if
necessary, we impose additional boundary conditions to ensure that it
does.  This completes the definition of the theory, and also
identifies the surface terms in $H_F$ as (at least some) of the
surface observables.

\item{(3)} Finally we ask if there are any other surface observables
that commute with the full Hamiltonian $H_F$. Since the boundary
conditions on the lapse and shift functions are not as stringent as in
the asymptotically flat case, we check to see if additional surface
observables may be found by introducing more freedom into the surface
parts of $H_F$ by replacing the lapse and shift by more general
functions. Once all the surface observables have been determined in
this way, we compute their algebra. This sets the stage for
quantization.

\end{description}

At the end of this procedure we have a well-defined theory, its
surface observables, and their algebra.  The full Hamiltonian of the
theory is functionally differentiable and satisfies a consistent
algebra.  On a solution, the full Hamiltonian may have non-vanishing
terms which are integrals on the boundary $\p\Sigma$.  These are the
surface observables.

Below we list the possible boundary conditions, and then follow the
rest of the procedure for two choices of spatial boundary conditions.
In the first case, all gauge parameters are set to zero on the spatial
boundary, while in the second the triad (and therefore the metric)
is fixed on the spatial boundary.

\subsection{Boundary conditions} 

We consider only local boundary conditions.  Instead of requiring that
integrals on the boundary vanish, we list the stronger conditions that
the integrand vanishes.  In this sense, our list is only a complete
list of { \em local} boundary conditions for gravity.  For certain
cases, such as where the boundary is a sphere, it is possible to
introduce global boundary conditions. Then, as in the asymptotically
flat case, one can impose parity conditions on the fields at the
boundary to make undesirable surface integrals vanish.

For completeness we first mention the conditions on the timelike
three-boundary $\p {\cal M}$, before listing the conditions on the spatial
two-boundary $\p\Sigma$.  The variation with respect to $A_a^i$ of the
first term in the $(3+1)$-action (\ref{3+1}) gives the surface term
\begin{displaymath}
\int_\Sigma d^3x\ E^{ai} \delta A_a^i\ |_{t_1}^{t_2}.
\end{displaymath}
This  can be made to vanish by  requiring $A_a^i$ to be 
fixed on $\p{\cal M}$, or by subtracting 
\begin{displaymath}
\int_{t_1}^{t_2} dt \int_\Sigma d^3x\ {d\over dt}(E^{ai}A_a^i)
\end{displaymath}
from the action and requiring $E^{ai}$ to be fixed on $\p{\cal M}$.

For the remainder of the paper we focus on the spatial two-boundary
$\p\Sigma$. The variation of the action (\ref{3+1}) contains the
variations of each constraint with respect to the phase space
variables. These can contribute a surface term to the full, finite
boundary Hamiltonian. Of course, in order to obtain the correct
initial value constraints for vacuum gravity, or BF theory, the gauge
parameters (Lagrange multipliers) have to be fixed on
$\p\Sigma$. Precisely how these parameters are fixed may depend, as we
discuss below, on what choices are made for the phase space variables
on $\p\Sigma$, and/or what gauge invariances on the boundary one would
like.

The following is the list of possibilities that gives functional
differentiability of the initial value constraints. Every mutually
consistent choice from this list defines a possible finite boundary
theory. We can of course change our starting point, and begin with a
($3+1$)-action that already has an arbitrary surface term, rather than
the action (\ref{3+1}). This obviously increases arbitrarily the
possibilities for defining theories in finite spatial regions. For
example, one could add a Chern-Simons term for the time like three-boundary
$R\times \p\Sigma$ as has been done by Smolin \cite{lee}. This leads
to source terms for the Gauss law constraint, and new possibilities
for boundary conditions.

\medskip
\noindent {\bf Gauss Constraint}
\medskip

The variation of the Gauss constraint is\footnote{We have overall
minus signs in all constraints because of the signs in the
$(3+1)$-action (\ref{3+1}).}
\bea
\delta G(\Lambda) &=& - \int_\Sigma d^3x\  \left[ \Lambda^i 
\epsilon^{ijk}
(\delta A_a^j E^{ak} + A_a^j \delta E^{ak})
-(\partial_a \Lambda^i) \delta E^{ai} 
+ \delta \Lambda^i D_a E^{ai} \right] \nonumber \\
&& -\int_{\partial\Sigma}d^2x\ n_a\Lambda^i \delta E^{ai},
\eea
where $n_a$ is the normal to the boundary two-surface
$\partial\Sigma$. The variation of the gauge parameter simply
yields the constraint. Functional differentiability with respect 
to the phase space variables requires vanishing integrand in 
the surface term, which leads to at least one of the following 
conditions:
\begin{description}
\item{(i)} Vanishing gauge transformations on the boundary
$$
\Lambda^i|_{\partial\Sigma}=0.
$$
\item{(ii)} Boundary conditions involving the triad:
\begin{description} \item{(a)} Fixed boundary ``area density''
$$
n_a \, \delta E^{ai}|_{\p\Sigma}=0.
$$
That this condition fixes the area density may be seen as follows.
Let $a^i = n_a E^{ai}$.  The surface area of the boundary is
$\int_{\p\Sigma} d^2x\ \sqrt{q} \equiv \int_{\p\Sigma}
d^2x\ \sqrt{ a^i \, a_i}$.  Fixing the area density means that $a^i
\delta a_i|_{\p\Sigma} = 0$ which is implied by the above
condition. 
\item{(b)} Fixed boundary triad  $\delta E^{ai}|_{\p\Sigma}=0$, 
or \item{(c)}  $\Lambda^i\delta E^{ai}|_{\p\Sigma}=0$.
\end{description}
\item{(iii)} Addition of the surface term\footnote{One is in fact free to 
add any multiple of this surface term. Functional differentiability then 
induces further conditions on fields and gauge parameters on the boundary.
Work is in progress on such special cases.}
$$
+\int_{\p\Sigma} d^2x\ n_a \left( \Lambda^i E^{ai} \right).
$$
\end{description}
These boundary conditions may be placed independently on different
disjoint parts of the boundary.  One could also take a combination of
cases, such as (i) and (ii).

\medskip
\noindent {\bf Diffeomorphism Constraint}
\medskip

The variation of the diffeomorphism constraint gives
\bea 
\label{vardiffeo}
\delta D(N^a) &=& - \int_\Sigma  d^3x\  
\left[  \delta E^{ai} { \cal L}_N A_a^i -
\delta A_a^i {\cal L}_N E^{ai} + \delta N^a {\cal D}_a \right] \nn
&+& \int_{\partial\Sigma} d^2x \ n_b \left[ N^a \delta(A_a^i E^{bi})
- N^b E^{ai} \delta A_a^i \right].
\eea
There are five choices which guarantee functional differentiability:
\begin{description}
\item{(i)} Vanishing diffeomorphisms on the boundary
\be
N^a|_{\p\Sigma} = 0.
\nonumber
\ee
This case is effectively the same as for manifolds without boundary.
The spatial diffeomorphism constraint in this case may be rewritten as
\begin{equation}
D(N) = - \int_\Sigma d^3x\ A_a^i {\cal L}_N E^{ai}.
\end{equation}
\item{(ii)} Addition of the boundary term
$$  
- \int_{\p\Sigma} d^2x\ n_b( N^a A_a^i E^{bi} )
$$
to $D(N)$, {\it and} restriction of the normal component of the shift 
on the boundary
$$
n_a N^a |_{\partial \Sigma} = 0.
$$
That is, the shift function on the boundary 2-surface $\p\Sigma$ must
be tangential to the boundary. For this choice, imposing functional
differentiability on $D(N)$ does not require that diffeomorphisms
vanish on the boundary.
\item{(iii)}  Addition of the same  boundary 
term as in (ii) {\it and} fixed connection on the boundary
$$
\delta A_a^i|_{\partial \Sigma}=0.
$$
\item{(iv)} Fixed triad (and hence metric) on the boundary, 
$$
\delta E^{ai}|_{\partial\Sigma}=0
$$
{\it and} addition of the boundary term
$$
- \int_{\partial\Sigma} d^2x\  n_b \left[ N^a A_a^i E^{bi}
- N^b E^{ai}  A_a^i \right],
$$
with the shift function free on the boundary.
\item{(v)} Fixed fields on the boundary 
$$
\delta E^{ai}|_{\partial\Sigma}=\delta
A_a^i|_{\partial\Sigma}=0.
$$
\end{description}

\medskip
\noindent {\bf Hamiltonian Constraint}
\medskip

The variation of the Hamiltonian constraint (\ref{ham})
is 
\bea
\delta H(N) &=& -\int_\Sigma d^3x\ 2\epsilon^{ijk}
\left[\ (N E^{bj}F_{ab}^k)\delta E^{ai}
+ (\epsilon^{klm} N E^{ai}E^{bj}A_b^m)\delta A_a^l
- \p_a (N E^{ai} E^{bj}) \delta A_b^k \right. \nn  
&& \left. +\ { 1 \over 2}  E^{ai} E^{bj} F_{ab}^k \delta N \ \right]  
 -\int_{\p\Sigma} d^2x \ n_a \, 2 \, \epsilon^{ijk}
 N E^{ai} E^{bj} \delta A_b^k .
\eea
Functional differentiability  requires at least one of the following:
\begin{description}
\item{(i)} Vanishing lapse on the boundary 
$$
N|_{\partial \Sigma} = 0.
$$
This eliminates the possibility of having a boundary Hamiltonian, and
hence dynamics and quasi-local energy.  However, it may be appropriate
for spacetimes containing a bifurcate Killing horizon.
\item{(ii)} The triad satisfies
$$
n_a E^{ai}|_{\partial \Sigma}  = 0
$$
which restricts the metric on the boundary to be tangential. 
This requires the spatial 3-metric to be degenerate on
the boundary. 
\item{(iii)} Boundary conditions involving the connection: 
\begin{description} 
\item{(a)} The variation of the tangential part of the connection 
vanishes on the boundary
$$
n_{[a} \delta A_{b]}^i |_{\partial \Sigma} = 0.
$$ 
\item{(b)} Or, the connection's variation vanishes 
$ \delta A_{a}^i |_{\p\Sigma} = 0.$ 
\end{description} 
\item{(iv)} Addition of the surface term
$$
+ \int_{\partial\Sigma}d^2x\ 2 N \epsilon^{ijk} A_a^i E^{aj} \, n_b \, E^{bk}
$$
and 
$$
\delta E^{ai}|_{\partial\Sigma}=0.
$$
This fixes the boundary 2-metric.  The surface term leads to the
quasi-local energy \cite{BY} and becomes the usual ADM surface energy
in the asymptotically flat case.

\end{description}

Adding a cosmological constant term to the Hamiltonian constraint 
does not contribute any new surface terms to the variation because 
it does not contain any derivatives, therefore the above choices of
boundary conditions remain the same. 

The asymptotically flat case in the Ashtekar variables has been 
worked out \cite{TT}. The fall off conditions on the lapse and 
shift are the same as for the ADM variables (Sec. 1.1) while 
the fall off conditions on the phase space variables are 
\be
A_a^i = {a_a^i(\theta,\phi)\over r^2} + O({1\over r^3}), \ \ \ \  
E^{ai} = e^{ai} + {f^{ai}(\theta,\phi)\over r} + O({1\over r^2}),
\ee
where $a_a^i$ and $f^{ai}$ are functions on the sphere at infinity,  
and $e^{ai}$ is a dreibein such that $e^{ai}e^{bi}=\delta^{ab}$.

We now consider two specific cases of functionally differentiable 
actions from the above list, and continue with steps (2) and (3)
for each case. Any other case may be similarly treated. 

\subsection{The case $\Lambda^i|_{\p\Sigma}=N^a|_{\p\Sigma}
=N|_{\p\Sigma}=0$}

Perhaps the simplest choice of boundary conditions is the case for
which all gauge parameters vanish on the boundary. This corresponds to
case (i) for each of the constraints.  The action is exactly the same
as for the closed case (\ref{3+1}), and therefore the constraint
algebra is just as in Eqs. (\ref{cont1} $-$ \ref{cont2}).  The
Hamiltonian remains a linear combination of constraints; all the
surface integrals vanish identically.  There are no surface
observables which arise as surface terms in the action.

Turning to step (3) above, we ask if there are any surface
observables.  One might expect that the reduction in gauge freedom
should give many surface observables: As the phase space variables on
the boundary are completely unconstrained, all the gauge degrees of
freedom in the interior become true degrees of freedom on the
boundary.  This does indeed occur, but the reduction of the gauge
freedom does not correspond directly to new observables in each case.
Rather, as we now see, there are an infinite number of observables,
but not an infinite number for each gauge parameter.

To find the explicit form of the observables, consider the 
functionals 
\bea
{ \cal O}_G(\lambda) &=& \int_\Sigma d^3x\ E^{ai} D_a \lambda^i; 
\label{gaussobs} \\
{ \cal O}_D(M) &=& \int_\Sigma d^3x\ A_a^i { \cal L}_M E^{ai}; \\ {
\cal O}_H(L) &=& \int_\Sigma d^3x\  \epsilon^{ijk} \left[ - 2 A_b^k \p_a
\left( L E^{ai} E^{bj} \right) + L E^{ai} E^{bj} \epsilon^{klm} A_a^l
A_b^m \right], \label{hamobs} 
\eea
where $\lambda^i$, $M^a$, and $L$ are (at this stage) arbitrary, and
{\it unconnected} with the gauge parameters $\Lambda^i$, $N^a$, and
$N$.  These functionals are obtained by integrating the constraints by
parts, discarding the surface terms, and replacing the gauge
parameters with the functions $\lambda$ $M^a$ and $L$. This approach
was followed by Balachandran, Chandar, and Momen in Ref. \cite{BCM}.
Since $E^{ai}$ and $A_a^i$ are free on the boundary, functional
differentiability is guaranteed if we require $L|_{\p\Sigma}=0$ and
$n_aM^a|_{\p\Sigma}=0$, leaving $\lambda^i$ arbitrary.  It is
important to note that functional differentiability eliminates ${\cal
O}_H$ as an observable.  The remaining functionals are surface
observables in that they are weakly equal to surface integrals
\bea
{ \cal O}_G (\lambda) &\approx & 
- \int_{\p\Sigma} d^2x\ n_a \lambda^i E^{ai} \\
{ \cal O}_D (M) &\approx & 
\int_{\p\Sigma} d^2x\ n_b E^{bi} M^a A_a^i.
\eea

It is easy to see that the non-zero ${\cal O}_G$ and ${\cal O}_D$ have
weakly vanishing Poisson brackets with the constraints; any possible
surface terms in their Poisson brackets with the constraints vanish
because the gauge parameters $\Lambda^i$, $N^a$, and $N$ vanish on the
boundary.

Given the definitions of the observables, the algebra\footnote{It is 
possible to show in general that the Poisson brackets of two 
functionals is functionally differentiable \cite{brownh}} is the 
expected one 
\bea
\left\{ {\cal O}_G(\lambda), {\cal O}_G(\mu) \right\} &=&
{\cal O}_G(\lambda \times \mu) ;  \\
\left\{ {\cal O}_D(M), {\cal O}_D(P) \right\} &=& {\cal O}_D \left(
\left[ M, P \right] \right) ; \\
\left\{ {\cal O}_G(\lambda), {\cal O}_D(M) \right\} &=&
- {\cal O}_G \left( { \cal L}_M \lambda \right).
\eea

Thus, we see that restricting the gauge freedom on the boundary
generates surface observables. However, as the case of ${\cal O}_H$
above shows, there need not be any direct correspondence between
reducing gauge degrees of freedom on the boundary and increasing the
number of boundary observables. The connection is more subtle; the new
degrees of freedom give more observables for the kinematic
constraints, but not for the Hamiltonian constraint.

\subsection{Fixed boundary metric}

The case of fixed triad on the spatial boundary $\delta
E^{ai}|_{\p\Sigma}=0$, and hence fixed boundary metric, is a more
interesting case. It has been studied before, although not entirely
along the lines we follow. Brown and York studied this case starting
from the standard metric action \cite{BY}, and gave definitions for
quasi-local quantities associated with the finite boundary.  Lau
performed an analysis similar for the fixed metric case in the new
variables \cite {SL}.  These works begin with the covariant action
rather than the (3+1)-action for the spatially closed case, and do not 
exhibit an algebra of surface observables.

Fixed boundary metric means case (ii) for the Gauss law and case (iv)
for the Hamiltonian constraint, but more than one possibility for the
diffeomorphism constraint.  The possible diffeomorphism cases are (i),
(ii) and (iv). Among these, we consider (ii) because it gives the
minimal restriction on the shift function, as well as a well-defined
algebra.  This gives the (3+1)-action
\bea
\label{fixedmetric}
S[E^{ai}, A_a^i; \Lambda^i, N^a, N] &=& 
\int dt\int d^3x\ [\ E^{ai}\dot{A}_a^i - N{\cal H} - 
N^a{\cal D}_a - \Lambda^i G^i\ ] \nn
&& + \int dt \int_{\p\Sigma}d^2x\ (2 N n_b 
\epsilon_{ijk}A_a^i E^{aj}E^{bk}) \nn
&& - \int dt \int_{\p\Sigma}d^2x\ (n_b N^a A_a^i E^{bi}),
\eea
where $N^a$ must be tangential to the boundary at the boundary.
Although we started with only the condition of fixed metric on the
boundary, the additional condition $n_a N^a|_{\p\Sigma}=0$ was induced
(by choice (ii) for diffeomorphisms). In general additional conditions
on the boundary may be induced by the Hamiltonian algebra and by
requiring the boundary conditions to be preserved in time.

Although this action represents a well-defined variational principle,
we are still free to add to it a surface term which is a function of
the fixed boundary data. This is an ambiguity in any variational
principle. For gravity, this freedom has been utilized \cite{BY,HH} to
normalize the values of the various surface observables relative to a
reference solution. This is done by subtracting the action of the
reference solution from the action of the solution of interest.  Such
normalizations may be necessary in order to avoid divergences of the
action, as in the asymptotically flat case, where integrations are
over all space.  Here we consider finite spatial regions so the action
(\ref{fixedmetric}) is well-defined and divergence-free as it stands.

The full Hamiltonian $H_F$ is a linear combination of constraints plus 
surface terms, and is identified from Eq. (\ref{fixedmetric}),  
\bea 
H_F[E^{ai}, A_a^i; \Lambda^i, N^a, N] & = & \int_\Sigma d^3x\ \left[
 N{\cal H} + N^a{\cal D}_a + \Lambda^i G^i \right] \nn
&& +\int_{\p\Sigma} d^2x\  n_b  
\left[ 2 N  \epsilon_{ijk}A_a^i E^{aj}E^{bk}
- N^a A_a^i E^{bi} \right].
\eea
Denoting the Hamiltonian constraint plus its corresponding surface
term by $H'$, and the diffeomorphism constraint plus its surface term
by $C$, the algebra of the full Hamiltonian contains
\bea
\{ G(\Lambda), G(\Omega) \} & = & G(\Lambda \times \Omega) 
+ \int _{\p\Sigma}d^2x\  n_cE^{ci}(\Lambda \times \Omega)^i, \\
\{ H'(M), H'(N) \} & = & -4C(K) + G (A_a K^a) 
- \int_{\p\Sigma} d^2x\ (n_aE^{ai})(A_b^iK^b),  
\eea
where $ K^a := E^{ai}E^{bi}(M\p_bN - N\p_bM) $. Similar surface terms
also arise in the Poisson brackets $\{ G(\Lambda), H'(N) \}$ and $\{
G(\Lambda), C(M) \}$.  All such surface terms ought to vanish in order
to have an anomaly free algebra. This may be accomplished by
requiring the lapse functions to be constant on the boundary and the
Gauss parameters to vanish on the boundary.  No additional constraint
on the shift function is required, (other than the already imposed
$n_aN^a|_{\p\Sigma}=0$).
 
For consistency it is also necessary that our boundary conditions be
preserved under evolution. This leads to further conditions. The piece
of $H_F$ which generates non-trivial evolution of $E^{ai}$ is $H'$,
(the Hamiltonian constraint plus its surface term). This leads to the
condition
\be
\dot{E}^{ai}|_{\p\Sigma} = 
 2N\epsilon^{ijk}D_b(E^{aj}E^{bk})|_{\p\Sigma} = 0. 
\label{metcond}
\ee
The simplest solution of this is to require that the lapse $N$ vanish
on the boundary.  This is rather limiting, however, because it means
that the quasi-local energy observable vanishes.  The only other
possibility is that the fixed boundary dreibein satisfy
Eq. (\ref{metcond}).  We choose the latter possibility $-$ the boundary
metric is required to be static.  We note that no further conditions
are necessary, in particular, the connection $A_a^i$ is free to vary
on the boundary with no consequences for functional differentiability.
In summary, the theory is  defined with the following conditions
\be
\delta E^{ai}|_{\p\Sigma} = 0,\ \ \ \  \dot{E}^{ai}|_{\p\Sigma} = 0,
\ee
\be
n_a N^a|_{\p\Sigma} = 0, \ \ \ \ \Lambda|_{\p\Sigma} =0, \ \ \ \ 
\p_a N|_{\p\Sigma} = 0.
\ee

The surface observables are just the surface terms in the action in
Eq. (\ref{fixedmetric}), with the lapse $N$ fixed to be constant. We
note that while there is only one quasi-local energy observable
\footnote{While this is another derivation, the surface term
(\ref{qle}) {\it is} the same as the quasi-local energy in
Ref. \cite{BY}, where it arises by varying the action with respect to
the boundary lapse function. Setting the (constant) lapse here to one
ensures that the normalizations are the same.}
\bea
\label{qle} 
 {\cal O}_H (N)  & = & \int_{\Sigma} d^3x \ \epsilon^{ijk} \left[ 
2 \p_b \left( N E^{ai} E^{bj} \right) A_a^k - \epsilon^{klm}
E^{ai} E^{bj} A_a^l A_b^m \right] \nn
& \approx & 2 N \int_{\partial\Sigma} d^2x\ (n_b \epsilon_{ijk} A_a^i 
E^{aj}E^{bk}), 
\eea
there are an infinite number of ``momentum'' observables 
\be
{\cal O}_D (N^a) = \int_{\Sigma} d^3x\  E^{ai} {\cal L}_N A_a^i 
\approx  \int_{\partial\Sigma} d^2x\ (n_b N^a A_a^i E^{bi})
\ee
parameterized by vector fields $N^a$ subject to
$n_aN^a|_{\p\Sigma}=0$. These are the generalization of the ADM
momentum and angular momentum for finite boundary.

As in the last section, we can ask if there are any other surface
observables defined like those of Eq. (\ref{gaussobs}-\ref{hamobs}).
One might think that there should be an infinite number of Gauss
observables as in Eq. (\ref{gaussobs}) because the gauge parameters
$\Lambda^i$ vanish on the boundary here (just as in last
subsection). However the algebra $\{H_F, {\cal O}_G \}$ contains the
piece $\{C(N), {\cal O}_G(\lambda) \}$ which weakly equals a surface
term unless $\lambda^i|_{\p\Sigma}= 0$. Thus, there are no surface 
observables other than the two above. 

The algebra of surface observables is {\it necessarily} the same as
the algebra of $H_F$ with itself. Indeed, the addition of boundary
terms may be viewed as accomplishing nothing but the functional 
differentiability of the constraints.
\bea
\left\{ { \cal O}_D (N), { \cal O}_D (M) \right\} &=& - { \cal O}_D 
\left( {\cal L}_N M \right) \nn
\left\{ { \cal O}_H (N), { \cal O}_H (M) \right\} &=& { \cal O}_D (K)\\ 
\left\{ { \cal O}_H (N), { \cal O}_D (M) \right\} &=& { \cal O}_H 
\left( {\cal L}_N M \right) \nn
\eea
where $K$ is given after Eq. (\ref{cont2}).

A comparison of the results of this and the last subsection shows 
that all surface observables are contained in the full Hamiltonian,  
except for the case (Sec. 2.2)
$$
\Lambda|_{\p\Sigma} = N^a|_{\p\Sigma} =
\p_aN|_{\p\Sigma} = 0.
$$

\section{BF theory in a bounded region}

We now turn to BF theory and apply the same procedure. The
topological BF theory in four dimensions has action
\be 
S = \int_M  {\rm Tr} \left[ B \wedge F + { \alpha \over 2 } 
B \wedge B
\right] 
\ee
where $F(A)$ is the curvature of a Yang-Mills gauge field and $B$ is a
Lie algebra valued two-form. We consider the case of gauge group
SU(2), and four-manifold $M=\Sigma\times R$, in which space $\Sigma$
has a boundary. The (3+1)- decomposition of this action leads to the
the phase space variables $A_a^i$, $E^{ai}=\epsilon^{abc}B_{bc}^i$,
and the first class constraints

\bea 
G^i\equiv D_aE^{ai} =0;\\
f^{ai}\equiv\epsilon^{abc} F_{bc}^i + { \alpha } E^{ai} = 0.
\label{f+e}
\eea 
On a three-manifold without boundary with $\alpha=0$, the theory has
two sets of Dirac observables\cite{vh}. One set depends on loops, and
the other on loops and closed two surfaces in $\Sigma$. The first is
the trace of the holonomy of $A_a^i$ based on loops $\gamma$,
$T^0[A](\gamma)= {\rm Tr}U_\gamma[A]$, and the second set is
\be
T^1[A,E](\gamma,S) = \int_S d^2\sigma\ n_a
{\rm Tr}[E^{a}(\sigma)U_\gamma(\sigma,\sigma)],
\ee 
where $n_a$ is the unit normal to the surface $S$, and $\sigma$ is the
basepoint of the loops $\gamma$. These are obviously invariant under
the Gauss constraint and a calculation shows that they are also
invariant under the second constraint. On the constraint surface,
these observables capture information about non-contractible loops and
closed two-surfaces in $\Sigma$. For example, for $\Sigma = S^1\times
S^2$, there is one observable of each type on the constraint surface.

We would like to find what additional observables, other than the
above bulk ones, arise when $\Sigma$ has boundary.\footnote{Boundary
observables and their quantization for Abelian BF theory has been
extensively discussed in a series of papers by Balachandran et.al
\cite{bal} } 

We therefore follow the procedure for gravity outlined in the previous
section. Since the second term in the action does not contain spatial
derivatives, the following applies to both zero and non-zero
cosmological constant.

\subsection{Boundary conditions}

The functional differentiability conditions for the Gauss law are 
as already outlined above in section 2.1. The constraint (\ref{f+e})
with gauge parameter $V_a^i$ is
\be
\label{bfconst}
F(V) = - \int d^3x\ V_c^i \left( \epsilon^{abc} F_{ab}^i 
+ \alpha E^{ai} \right). 
\ee
Its variation is 
\begin{equation}
\delta F(V) = 2 \int_{\partial \Sigma} d^2x\    
\epsilon^{abc}n_a \delta A_b^i
V_c^i - \int_\Sigma d^3x\  \left[ 2 \epsilon^{abc}
\left( \partial_a V_c^i \delta A_b^i
- \epsilon^{ijk} V_c^i A_a^j \delta A_b^k \right) 
+ \alpha V_a^i \delta E^{ai} \right].
\end{equation}
Functional differentiability leads to the following choices: 
\begin{description}
\item{(i)}   
$$
 \epsilon^{abc} n_b  V_c^i |_{\partial \Sigma} = 0;
$$
\item{(ii)} 
$$
\ V_a^i|_{\partial \Sigma} = 0; 
$$
\item{(iii)}  Addition of  the boundary term
$$
+ 2 \int_{\partial \Sigma}d^2x\ \epsilon^{abc}n_a A_b^i V_c^i. 
$$
\item{(iv)} Conditions involving the connection: (a) 
$$
\delta A_a^i |_{\p\Sigma} =0, 
$$
or (b) 
$$
n_{[b}\delta A_{a]}^i |_{\p\Sigma} =0,
$$
\end{description}

Perhaps the most interesting case is to keep the gauge transformations
unrestricted on the boundary. We therefore consider cases (iii) for
both the Gauss constraint and the BF theory constraint
(\ref{bfconst}). The (3+1)-action for this case is
\bea
S[E^{ai}, A_a^i; \Lambda^i, V_a^i] =
\int dt \int_{\Sigma} d^3x\ [ E^{ai}\dot{A}_a^i - V_a^if^{ai}
-\Lambda^iG^i] \nn 
+ \int dt \int_{\p\Sigma}d^2x  n_a \left[ 2 \epsilon^{abc} A_b^i V_c^i 
+  E^{ai}\Lambda^i \right]. 
\eea
>From this we identify the Hamiltonian 
\bea
H_F [E^{ai}, A_a^i ; \Lambda^i, V_a^i ] &=& \int_\Sigma d^3x\
[ V_a^if^{ai}
-\Lambda^iG^i] + \int_{\p\Sigma}d^2x\  n_a \left[ 2 \epsilon^{abc} A_b^i V_c^i 
+  E^{ai}\Lambda^i \right] \nn
&=& \int_\Sigma d^3x\ [2 \epsilon^{abc} D_b V_a^i A_c^i - \alpha V_a^i E^{ai} 
+ E^{ai} D_a \Lambda^i]
\label{bffull}
\eea
As before, for any specific choice of boundary conditions, we must
calculate the algebra of the full Hamiltonian with itself.  If the
algebra does not close then we need further conditions on the gauge
parameters $\Lambda^i$ and $V_a^i$.  Denoting $H_F =
f'(V)+G'(\Lambda)$ where $f'$ and $G'$ are the constraints plus their
corresponding surface terms, we find that the algebra $\{
H_F(V,\Lambda), H_F(W,\mu)\}$ contains
\bea 
\{ G'(\Lambda), G'(\mu) \} &=& G'(\Lambda \times \mu),\\
\{ G'(\Lambda), f'(V) \} &=& 2f'(\Lambda\times V) 
+ 2\int_{\p\Sigma}d^2x\ \epsilon^{abc}n_a\Lambda^i\p_bV_c^i, 
\label{gaussbf} \\
\{ f'(V), f'(W) \} &=&  \alpha \int_{\p\Sigma}d^2x\  
n_a \epsilon^{abc} W_b^i V_c^i 
\label{bfbf}
\eea
where $\Lambda\times V_c = \epsilon^{ijk} \Lambda^j V^k_c$.  We do not
want $V_a^i$ to vanish on the boundary because this would give a
vanishing surface observable. An alternative choice is to require
$V_a^i$ to be curl free on the boundary so that the surface term in 
Eq. (\ref{gaussbf}) vanishes. Finally, for $\alpha\ne 0$,
we require the surface term in Eq.
(\ref{bfbf}) to vanish. The least restrictive way to ensure this is
to require all the field $V_c^i$ etc. to tend to a fixed value on the
boundary. This completes the list of conditions.

As for gravity, on the constraint surface, the bulk parts of the full
Hamiltonian vanish leaving the surface terms as the surface
observables
\bea 
{ \cal O}_G(\Lambda) & \approx  &-\int_{\p\Sigma} d^2x\ 
n_a \Lambda^iE^{ai} \\
{ \cal O}_F(V)& \approx & -2 \int_{\p\Sigma} d^2x\ \epsilon^{abc} 
n_aA_b^iV_c^i.
\eea 
These observables are parameterized by $\Lambda^i$ and $V_a^i$, and
therefore are infinite in number. Continuing to step (3), we find that
no more observables are obtained by generalizing the parameterizing
functions $\lambda^i$ and $V_c^i$ as in Sec. 2.2.

The observable algebra is
\bea
\left\{ { \cal O}_G (\Lambda), { \cal O}_G (\Omega) \right\} &=& { \cal O}_G
(\Lambda \times \Omega) \nn
\left\{ { \cal O}_F (V), { \cal O}_F (W) \right\} &=& 0 \\
\left\{ { \cal O}_F (V), { \cal O}_G (\lambda) \right\} &=& { \cal O}_F
( \Lambda \times V).
\eea
Finally, since the bulk observables for vanishing cosmological
constant ($\alpha=0$) based on loops and surfaces commute with the
full Hamiltonian, the bulk observables commute with the surface
observables.

\subsection{Quantization}

In the connection representation, BF theory with non-zero 
cosmological constant, $\alpha$, has
a unique solution \cite{hor}. The quantum constraints are
\bea 
D_a{\delta\over \delta A_a^i}\psi[A] &=& 0; \\
(\epsilon^{abc} F_{bc}^i +  \alpha 
{\delta\over \delta A_a^i})\psi[A] &=& 0.
\eea
A regularization is not necessary here because the equations are
linear in the momenta.  The unique solution, for $\Sigma$ without
boundary, is the Chern-Simons state
\be 
 \psi[A] = {\rm exp}\left[-{1\over \alpha} 
\int_\Sigma {\rm Tr}(A\wedge dA + {2\over 3} A\wedge A\wedge A)
\right].
\label{cs}
\ee
For spaces with boundary, the variation of the Chern-Simons state is
well defined only if the connection is fixed on the boundary. With
this additional condition, although there are an infinite number of
boundary observables, the Hilbert space has only one state. Its
eigenvalue depends on the surface terms in the Hamiltonian. This
example shows that an infinite number of boundary observables does not
preclude a finite dimensional Hilbert space for the bounded system. In
fact, if we take the case (iv) for the BF theory constraint we have
only the Gauss observable and its algebra. Since this is just the
angular momentum algebra, quantization can give a finite dimensional
state space.  Thus this theory, though meager in content, is not
inconsistent with the holographic hypothesis.

\section{Discussion}

Beginning with the Hamiltonian action, we studied Einstein gravity and
BF theory in finite spatial regions.  The action had to be augmented
by surface terms in order to generate the correct equations of motion.
After giving a complete list of local boundary conditions, we
identified surface observables and computed their algebra. These
observables naturally arose from the surface terms added to the
action.  We noted that additional surface observables may be generated
in some cases (as in Sec. 2.2) by replacing the gauge parameters with
more general functions.

The procedure given here is similar in spirit to that of Regge and
Teitelboim \cite{rt}. Imposing functional differentiability on the
(3+1)-action results, in most cases, in the addition of surface terms
to the action.  So when a boundary is present, there is non-vanishing
full Hamiltonian which is a linear combination of constraints plus
surface terms. Evaluated on a solution, this Hamiltonian gives the
non-vanishing surface observables.  All conditions on the functions
parameterizing the observables are derived by requiring that the
algebra of the full Hamiltonian remain anomaly-free. In the interior,
this Poisson bracket gives the algebra of constraints. When restricted
to the boundary, it gives the algebra of surface observables. Also,
except for the case studied in Sec. 2.2, there are no surface
observables other than those which already make up the full
Hamiltonian - as we saw any attempt to ``generalize'' the parameters
in the surface terms leads to undesirable surface terms in the
algebra.

Although our discussion was restricted to the case of a single finite
boundary, other cases are easy to incorporate. For example, for a
static black hole, one would have both an inner boundary at the
horizon and  an asymptotic boundary.  One could choose the
boundary conditions given in Sec. 2.2 on the horizon and the standard
conditions in the asymptotic region \cite{TT}.  For a black hole in a
bounded region, one would need to augment the procedure to include an
additional finite inner boundary.  On this boundary one could use the
conditions given in Sec. 2.3.

More generally, this procedure could be used for relating an observed
spacetime to an observer spacetime.  This ``relative state formalism''
is an extension of the study of asymptotic spacetimes and closely related
to methods of topological field theory \cite{jb}.  The
system may be expressed as a known classical solution matched with a
gravitational system of interest.  By cutting the (compact) space
$\Sigma$ in two pieces, say $\sigma_1$ and $\sigma_2$, we could
express the space as
$$
\Sigma = \sigma_1 \cup \sigma_2.
$$
The full Hamiltonian of the theory would split into full Hamiltonians
on each subspace
$$ 
H_\Sigma = H_{\sigma_1} + H_{\sigma_2}
$$ 
with, typically, the same bulk pieces and surface terms of opposite
sign in the individual Hamiltonians $H_{\sigma_i}$.  States in one
region are then expressed relative to the states
in the other region.  Such a formalism might provide a tractable
approach to quantization.

Although our primary focus is classical, we comment briefly on
quantization in the connection representation.  For general relativity
with a cosmological constant $\alpha$, it is known that the
exponential of the Chern-Simons integral (\ref{cs}) is a formal
solution to all the quantum constraints \cite{kodama}. In the fixed
metric case, the action of the full Hamiltonian on this state gives a
non-zero answer determined by the action of the surface terms. This
state is formally an eigenstate of the full Hamiltonian with
eigenvalue
\be 
 {1\over \alpha^2}\int_{\p\Sigma}d^2x\ 2 n_b \epsilon_{ijk}A_a^i B^{aj}B^{bk}
-{1\over \alpha} \int_{\p\Sigma}d^2x\ n_b N^a A_a^i B^{bi}, 
\ee
where $B^{ai} = \epsilon^{abc}F_{bc^i}$ is the restriction of the
magnetic field to the boundary. This state is the {\it only} solution
of the constraints of the topological BF theory. Thus this solution,
though an eigenstate of the Hamiltonian, corresponds to a
topological sector of general relativity.

It is interesting to see how our results can be made a part of the
ongoing developments in non-perturbative quantum gravity, quite apart
from extending the classical case to null inner boundaries.  For
example, given a system such as a black hole, one would select
appropriate boundary conditions, and find the action and the surface
observables.  From a suitable Poisson algebra of observables and one
state, the GNS construction provides a way of finding an inner
product, Hilbert space, and representation of the observable algebra.
A not unrelated direction is to fully explore the relative state
formalism by introducing a dynamical boundary, in which the theory is
essentially the bulk theory, find observables, and again use the GNS
construction to built a quantum theory.  In this manner, these
suggestions provide a basis for exploring the role of surface theories
in the context of gravitational entropy and the holographic
hypothesis.

\section*{ACKNOWLEDGEMENTS}

We thank the Mathematics Department of the University of New Brunswick
for hospitality where some of this work was done.  We also thank Chris
Beetle, Roumen Borissov, Domenico Giulini and Lee Smolin for comments
on the manuscript.  This work was supported, in part, by NSF grant
number PHY-9514240 to the Pennsylvania State University.

\end{document}